\documentclass[prx,reprint,superscriptaddress,twocolumn,showkeys]{revtex4-1}
\usepackage[utf8]{inputenc} 
\usepackage{amsmath}
\usepackage{braket}
\usepackage{qcircuit}
\usepackage{graphicx}
\usepackage{pgfplots}
\pgfplotsset{compat=1.15}
\usepackage{csquotes}
\usepackage{hhline}
\usepackage{amssymb}

\usepackage{dcolumn}
\usepackage{tabularx}
\usepackage[colorlinks=true,linkcolor=blue,citecolor=blue,urlcolor=blue]{hyperref}
\usepackage{braket}
\usepackage{float}
\usepackage{enumerate}
\usepackage{listings}
\usepackage{color}
\usepackage{subfigure}
\usepackage{placeins}
\usepackage{natbib}
\definecolor{mygreen}{rgb}{0,0.6,0}
\definecolor{mygray}{rgb}{0.5,0.5,0.5}
\definecolor{mymauve}{rgb}{0.58,0,0.82}

\newcolumntype{C}{>{\centering\arraybackslash}X}

\begin{document}
\title{Generation and characterization of customized perfect Laguerre-Gaussian beams with arbitrary profiles}
\author{Chengyuan Wang}
\email{wcy199202@gmail.com}
\affiliation{Ministry of Education Key Laboratory for Nonequilibrium Synthesis and Modulation of Condensed Matter, Shaanxi Province Key Laboratory of Quantum Information and Quantum Optoelectronic Devices, School of Physics, Xi'an Jiaotong University, Xi'an 710049, China}
\affiliation{Two authors have equal contributions on this work.}
\author{Yun Chen}
\affiliation{Ministry of Education Key Laboratory for Nonequilibrium Synthesis and Modulation of Condensed Matter, Shaanxi Province Key Laboratory of Quantum Information and Quantum Optoelectronic Devices, School of Physics, Xi'an Jiaotong University, Xi'an 710049, China}
\affiliation{Department of Physics, Huzhou University, Huzhou 313000, China}
\affiliation{Two authors have equal contributions on this work.}
\author{Jinwen Wang}
\affiliation{Ministry of Education Key Laboratory for Nonequilibrium Synthesis and Modulation of Condensed Matter, Shaanxi Province Key Laboratory of Quantum Information and Quantum Optoelectronic Devices, School of Physics, Xi'an Jiaotong University, Xi'an 710049, China}
\author{Xin Yang}
\affiliation{Ministry of Education Key Laboratory for Nonequilibrium Synthesis and Modulation of Condensed Matter, Shaanxi Province Key Laboratory of Quantum Information and Quantum Optoelectronic Devices, School of Physics, Xi'an Jiaotong University, Xi'an 710049, China}
\author{Hong Gao}
\email{honggao@mail.xjtu.edu.cn}
\affiliation{Ministry of Education Key Laboratory for Nonequilibrium Synthesis and Modulation of Condensed Matter, Shaanxi Province Key Laboratory of Quantum Information and Quantum Optoelectronic Devices, School of Physics, Xi'an Jiaotong University, Xi'an 710049, China}

\author{Fuli Li}
\affiliation{Ministry of Education Key Laboratory for Nonequilibrium Synthesis and Modulation of Condensed Matter, Shaanxi Province Key Laboratory of Quantum Information and Quantum Optoelectronic Devices, School of Physics, Xi'an Jiaotong University, Xi'an 710049, China}

\begin{abstract}
We experimentally demonstrate the generation of customized  Laguerre-Gaussian (LG) beams whose intensity maxima are localized around any desired curves. The principle is to act with appropriate algebraic functions on the angular spectra of LG beams. We characterize the propagation properties of these beams and compare them with non-diffraction caustic beams possessing the same intensity profiles. The results manifest that the customized-LG beams can maintain their profiles during propagation and suffer less energy loss than the non-diffraction caustic beams, and hence are able to propagate a longer distance. Moreover, the customized-LG beam exhibits self-healing ability when parts of their bodies are blocked. This new structure beam has potential applications in areas such as optical communication, soliton routing and steering, optical tweezing, etc.
\end{abstract}

%\begin{keywords}{Quantum teleportation, Hypergraph states, IBM quantum experience, Quantum state tomography}\end{keywords}
\maketitle

\section{Introduction}
Structure lights with customized intensity, phase, and polarization distributions \cite{rubinsztein2016roadmap,forbes2021structured} have promoted the development of fundamental physics \cite{hansen2016singular}, optical trapping \cite{10.1117/1.AP.3.3.034001,10.1063/5.0013276,baumgartl2008optically}, imaging \cite{vettenburg2014light}, optical communications \cite{torres2012multiplexing,10.1063/5.0054885}, and light-matter interactions \cite{10.1116/5.0016007,ZhenKun134201,Fuqiang52304,wang2021efficient,Chen:21,wang2022experimental}, etc. Hermite–Gaussian (HG) beam\cite{wang2016generalised}, Laguerre–Gaussian (LG) beam \cite{PhysRevA.45.8185}, and Ince–Gaussian beam \cite{Bandres:ince,Yu:21} are three representative structure lights, which are the solutions of the paraxial wave equation under Cartesian, polar, and elliptical coordinates, respectively. These beams can remain their transverse profiles unchanged for about one Rayleigh length propagation, but suffer from an overall scaling of light spot due to diffraction. In 1987, by solving the Helmholtz equation in circular coordinates, Durnin proposed a well-known non-diffracting beam \cite{PhysRevLett.58.1499}, namely the Bessel beam, whose transverse structures neither deform nor expand over a significant propagation distance. However, an ideal Bessel beam with infinite energy is inaccessible practically. Thus in most realistic experiments, the Bessel beam is apodized to a Bessel-Gaussian (BG) beam \cite{McLaren:12,chu2015generating,zhi2023chip} that can be approximated as non-diffracting. Alternatively, some other forms of non-diffracting beams, e.g., Mathieu \cite{Gutierrez-Vega:00} and Weber beams \cite{Bandres:04}, can be obtained by extending the frame for solving the wave equation in elliptic and parabolic coordinates.

The above conventional non-diffracting beams have fixed, relatively simple transverse intensity distributions due to the limited classes of light fields, which in turn astricts their usefulness. To overcome this, several productive ways have been proposed \cite{Sanchez-Serrano:12,PhysRevLett.105.013902,Yan:21,wang2022experimental} to generate various types of non-diffracting or propagation-invariant beams, and most of them are based on spatial spectrum engineering techniques. It should be noted that the earliest related research can be traced back to the 1990s, in which the spiral-type beam in the form of arbitrary curves was put forward \cite{ABRAMOCHKIN1993336,ABRAMOCHKIN1996302} and later proved to be a class of caustic beam with propagation-invariant and resistance to perturbation \cite{Volyar:21}. As another representative example of customizing complex caustic beam, in 2020 Zannotti \emph{et al.} constructed arbitrary-shaped non-diffracting caustic beams through careful amplitude and phase modulation of the Bessel beam in Fourier space \cite{2020Shaping}, also known as the Bessel pencil method. This method utilizes the most localized propagation-invariant light spot, namely a 0th-order Bessel beam, as a `pencil' to draw the desired beam along a preset curve. Shortly thereafter, J. Mendoza-Hern\'{a}ndez theoretically proposed that by acting an algebraic function on an LG beam in Fourier space, a new structured beam similar to the non-diffracting caustic beam can also be generated \cite{Mendoza-Hernandez:21}, and importantly, such a beam can preserve its profile a longer propagation distance than the non-diffracting caustic beam. Nevertheless, this work does not extend to generating arbitrarily shaped beams and yearning for experimental realization.

In this paper, we experimentally generate customized-LG beams with any desired shapes. The rationale is based on tailoring the LG beams in Fourier space by algebraic functions \cite{Martinez-Castellanos:15,Mendoza-Hernandez:19,Mendoza-Hernandez:21}, which are constructed using the Bessel pencil method. To assess the propagation properties of the customized-LG beams, we also generate non-diffracting customized-BG beams \cite{2020Shaping} with the same transverse profiles for peer comparison. The experimental results show that the customized-LG beams can maintain their profiles during propagation and suffer less energy loss than the customized-BG beams, and hence can propagate a longer distance. Moreover, the self-healing ability of the customized-LG beam is also verified.

%%%===============================================%%%
\begin{figure*}[tb]    %此为跨栏图片写法，本杂志通栏宽度180mm,单栏宽度88mm，栏间距4mm
\vspace*{1mm}    %图片上空1mm
\centering    %图片居中
\includegraphics[width=13cm]{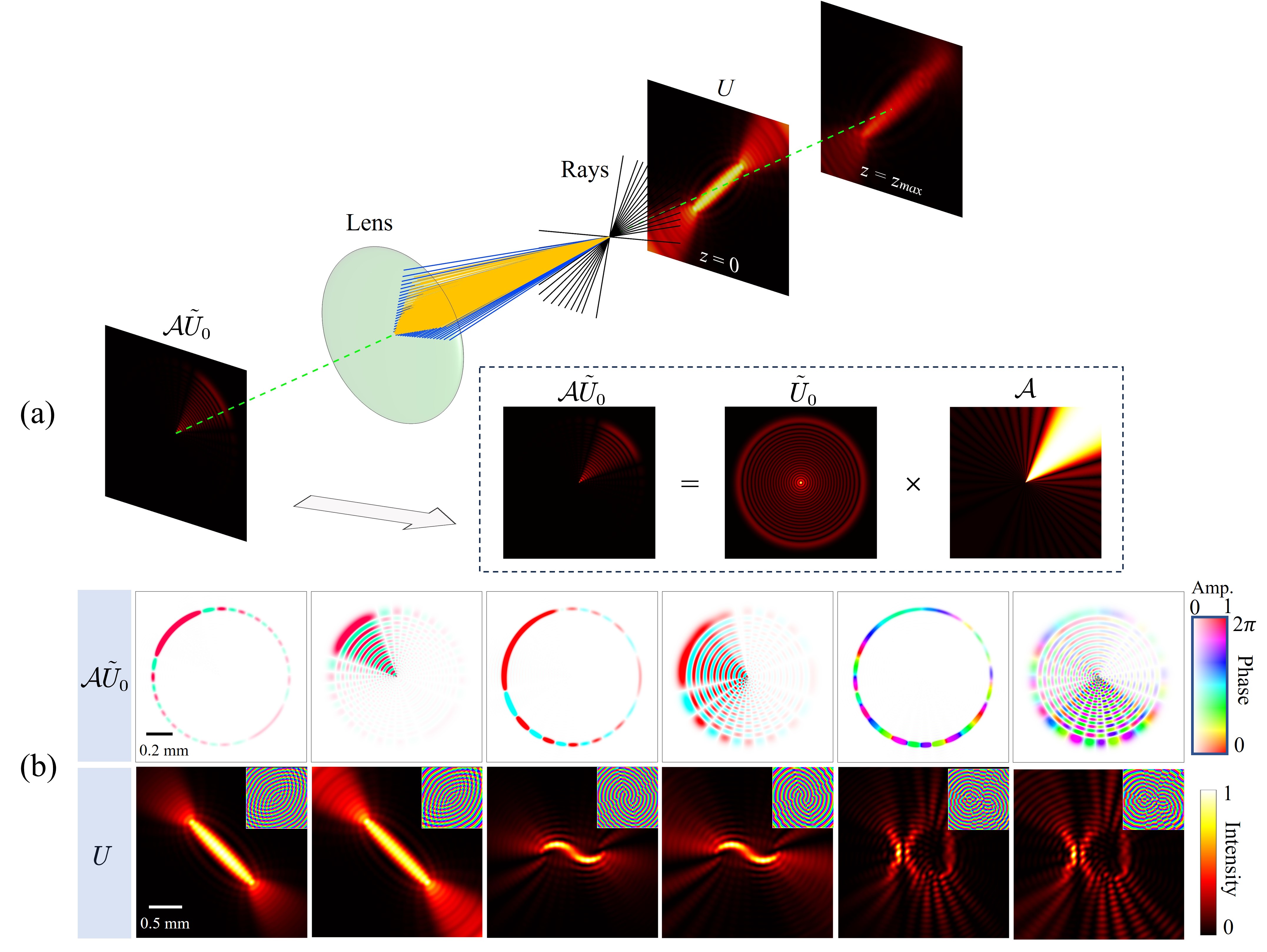}    %图片写法及名称格式
\\[1mm]    %图片与图说之间的距离，可修改，但应统一
\caption{
\textbf{(a)} Schematic diagram of generating customized-LG beams. \textbf{(b)} Intensity and phase distribution of modulated spatial spectrum $\mathcal{A} \tilde{U}_{0}$ (first row) and the corresponding customized beams $U$ (second row) with shapes of `sheet', `sinusoid', and `XJ', respectively. Note that two types of seed beams are used here, with columns 1, 3, and 5 corresponding to BG beam and columns 2, 4, and 6 corresponding to LG beam.)
\hfill{}
}\label{fig:1}
%\vspace*{0.5mm}    %图片下空0.5mm
\end{figure*}
%%%===============================================%%%

\section{Principles}\label{II}
%%%===============================================%%%
We first introduce the underlying principle. When a linear differential operator $\hat{\mathcal{A}}$ acts on an appropriate seed beam $U_{0}(r,z)$, a new customized beam $U(\mathbf{r},z)$ can be obtained with a complex transverse shape \cite{Martinez-Castellanos:15,Mendoza-Hernandez:21} or on-demand trajectory \cite{Yan:21,Cizmar:09}. Such a transformation process can be expressed as
%%%===============================================%%%
\begin{equation}\label{Eq:1}
U(\mathbf{r},z)=\hat{\mathcal{A}}U_{0}(r,z),
\end{equation}
%%%===============================================%%%
where $\mathbf{r}=(x,y)$ is the transverse coordinate. Using the Fourier transform $\mathcal{F}$ and inverse Fourier transform $\mathcal{F}^{-1}$, Eq. (\ref{Eq:1}) can be further written as
%%%===============================================%%%
\begin{equation}\label{Eq:2}
U=\mathcal{F}^{-1}\left\{\mathcal{F}\left\{\hat{\mathcal{A}} U_{0}\right\}\right\}=\mathcal{F}^{-1}\left\{\mathcal{A} \mathcal{F}\left\{U_{0}\right\}\right\}=\mathcal{F}^{-1}\left\{\mathcal{A} \tilde{U}_{0}\right\},
\end{equation}
%%%===============================================%%%
where $\mathcal{A}$ is an algebraic function corresponding to the operator $\hat{\mathcal{A}}$ in Fourier space \cite{Martinez-Castellanos:15}, $\tilde{U}_{0}$ is the Fourier transform of $U_{0}$ and can be optically obtained by a lens focusing process. The transformation process described in Eq. (\ref{Eq:2}) can be physically realized by the setup shown in Fig. \ref{fig:1}(a).

Since we focus on the seed of an LG beam and derive a collection of customized-LG beams based on the theory above, it is convenient to work with the expression of seed LG at the plane $z=0$
\begin{equation}\label{Eq:4}
L G U_{0}=C_{\mathrm{LG}} L_{p}^{|0|}\left[\frac{2 r^{2}}{\omega_{k t}^{2}(0)}\right] \exp \left[-\frac{r^{2}}{\omega_{k t}^{2}(0)}\right],
\end{equation}
where $C_{\mathrm{LG}}$ is a constant coefficient, $L_{p}^{|0|}$ is the Laguerre polynomial with $p$ being the radial order and azimuthal order $l=0$, $\omega_{k t}(0)=\omega_{0} / 2 \sqrt{2 N}$ with $\omega_{0}$ being the input Gaussian beam waist and $N=p+1/2$. It should be emphasized that when $p\gg 1$, the LG beam can be regarded as a quasi propagation-invariant beam comparable to the BG beam. Thereupon, to evaluate the propagation characteristic of the customized-LG beams, we also generate customized-BG beams with similar shapes, i.e. a conventional light field with non-diffraction properties, for comparison. In general, the seed BG beam can be expressed by
\begin{equation}\label{Eq:8}
BU_{0}=C_{\mathrm{B}} J_0\left(k_t r\right) \exp \left(-\frac{r^2}{\omega_{\mathrm{B}}^2}\right),
\end{equation}
where $C_{\mathrm{B}}$ is a constant, $J_0\left(.\right)$ is the zeroth order Bessel function of the first-kind, $k_t$ is the transverse component of the wave vector $k=2\pi/\lambda$, $\omega_{\mathrm{B}}$ is the waist of the BG beam \cite{Mendoza-Hernandez:15}. Note that, the requirements for an LG beam to behave similarly to a BG beam should be $w_{\mathrm{B}}=\sqrt{2 N}w_{\mathrm{0}}$ and $k_t=2\sqrt{2 N}/w_{\mathrm{0}}$ \cite{Mendoza-Hernandez:15}, then the two beams have the same maximum propagation distance $z_{max}=kw_{\mathrm{0}}^{2}/2$ within which their spatial structure does not undergo obvious deformation during propagation. Interestingly, these relations are also often used to produce the perfect LG beam whose spot size does not change with its orbital angular momentum \cite{Mendoza-Hernandez:20,doi:10.1063/5.0048741}.

%In this work take advantage of the above conditions to generate that is reasonable for peer comparison. 
The generation of several types of customized-LG beams, such as astroid-LG and deltoid-LG beams, have been theoretically proposed before \cite{Mendoza-Hernandez:21} and verified in our experiments (see the supplementary material for details). Extending it to customizing arbitrary-shaped LG beams is somewhat challenging, mainly because it is difficult to construct the corresponding algebraic function $\mathcal{A}$. Inspired by the formal similarity between the LG and BG beams, we found that the Bessel-pencil method could also be utilized to construct complex customized-LG beams. The algebraic function of such a beam can be constructed by

\begin{equation}\label{Eq:5}
\mathcal{A}=\int_{a}^{b} \exp \left[-\mathrm{i} k_{t} \mathbf{r}_{\mathrm{c}}(\tau) \cdot \mathbf{u}(\varphi)+\mathrm{i} \gamma_{\mathrm{B}}(\tau)\right] \mathrm{d} \tau,
\end{equation}
where $\mathbf{r}_{\mathrm{c}}=(x_{\mathrm{c}},y_{\mathrm{c}})$ is the transverse coordinate along the desired curve, $\mathbf{u}(\varphi)=(\mathrm{cos}\varphi,\mathrm{sin}\varphi)$ is a unit vector related to the azimuthal angle, and $\tau$ is the arc length of the curve, and $a/b$ is the head/tail of the curve. $\gamma_{\mathrm{B}}(\tau)$ is a phase term growing with the curve’s arc length $\tau$, which can be obtained by
\begin{equation}\label{Eq:6}
\gamma_{\mathrm{B}}(\tau)=k_{t} \int_{0}^{\tau}\left|\mathbf{r}_{\mathrm{c}}^{\prime}(s)\right| \mathrm{d} s.
\end{equation}

Based on the theory above we can obtain an arbitrary customized beam with an LG-seed or a BG-seed. For instance, we integrate Eq. (\ref{Eq:5}) along three designed curves, i.e., `sheet', `sinusoid', and letters `XJ', and obtain the corresponding spatial angular spectrum modulation, and then obtain three kinds of customized-LG and customized-BG beams by Fourier transform, as shown in Fig. \ref{fig:1}(b). For convenience, the sheet-shaped beam customized by LG or BG is abbreviated as LG/BG-sheet, and LG/BG-sine, LG/BG-XJ are named in the same way. The first row manifests that $\mathcal{A}$ acts as a phase and amplitude modulator on the angular spectrum, obtained by simulating the focusing process through a lens with $f=20$ cm, of the two seed beams, whose parameters are set as $p=20$, $w_{\mathrm{0}}=0.2$ mm, $\omega_{B}=\sqrt{2(p+1/2)}\omega_{0}=1.3$ mm and $k_{t}=2\sqrt{2(p+1/2)}/\omega_{0}=6.4\times 10^{4}$ m$^{-1}$,  such that they have similar intensity distribution and the same maximum propagation distance of  $z_{max}=kw_{\mathrm{0}}^{2}/2=16$ cm. The second row of Fig. \ref{fig:1}(b) exhibits the intensity and phase distributions of the corresponding customized-BG and customized-LG beams. It can be observed that using the same algebraic function, the shapes of customized-BG beam and customized-LG beam are almost the same with the peripheral sideband ignored, and their intensity maxima are indeed localized around the preset curve. Essentially, the customized beams can be approximately regarded as an ordered coherent superposition of the seed beams, and therefore possess some properties, such as propagation invariance and self-healing properties, of the corresponding seed beams, which we will focus on verifying experimentally in the following.

\begin{figure}[tb]    %此为跨栏图片写法，本杂志通栏宽度180mm,单栏宽度88mm，栏间距4mm
\vspace*{1mm}    %图片上空1mm
\centering    %图片居中
\includegraphics[width=8cm]{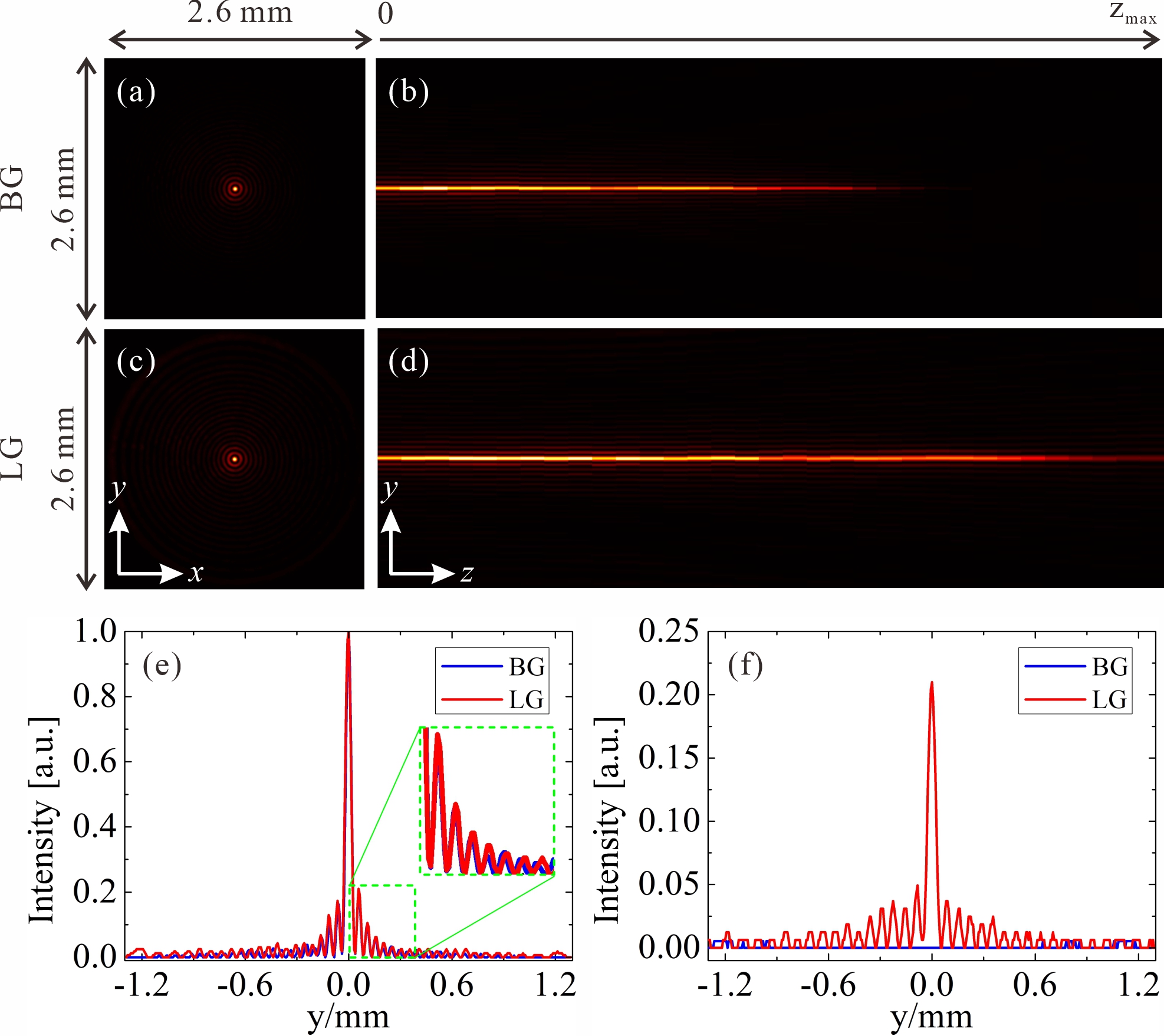}    %图片写法及名称格式
\\[1mm]    %图片与图说之间的距离，可修改，但应统一
\caption{
Intensity distributions of the experimentally generated \textbf{(a)} BG and  \textbf{(b)} LG beams. \textbf{(b)} and \textbf{(d)} are the beams' sectional propagation trajectories from $z=0$ to $z=z_{max}=16$ cm, and \textbf{(e)} and \textbf{(f)} are the intensity profiles of the two beams in the central region at propagation distances of $z=0$ and $z=16$ cm, respectively. )
\hfill{}
}\label{fig:2}
%\vspace*{0.5mm}    %图片下空0.5mm
\end{figure}

\section{Experimental results and discussions}
%%%===============================================%%%
The detailed experimental configuration is given in the supplementary material. Given that the seed beam determines the propagation characteristics of customized beams, the propagation evolution of the seed beams in free space are firstly investigated. The parameters used here are completely consistent with those in Fig. \ref{fig:1}.

Figure \ref{fig:2}(a) and \ref{fig:2}(c) show the intensity distributions of the seed BG and LG beams, \ref{fig:2}(b) and \ref{fig:2}(d) are the beams' sectional propagation trajectories from $z=0$ to $z=z_{max}$, and \ref{fig:2}(e) and \ref{fig:2}(f) are the intensity profiles of the two beams in the central region at propagation distances of $z=0$ and $z=z_{max}$, respectively. As can be seen that the two beams have nearly the same intensity distribution at $z=0$ as expected. During propagation, the BG beam maintains its overall profile but its intensity attenuates quickly whereas the LG beam maintains its intensity profile much better but suffers slight broadening, e.g., at $z=z_{max}$ the BG is almost dissipated while the intensity maximum of the LG beam keeps to 21\% of its original. This striking comparison indicates that compared with the customized-BG beam, the customized-LG beams suffer less intensity dissipation during long-distance propagation.

\noindent
%%%%%%%%%%%%%%%%%%%%%%%%%%%%%%%%%%%%%%%%
\begin{figure*}[tb]    %此为跨栏图片写法，本杂志通栏宽度180mm,单栏宽度88mm，栏间距4mm
\vspace*{1mm}    %图片上空1mm
\centering    %图片居中
\includegraphics[width=13cm]{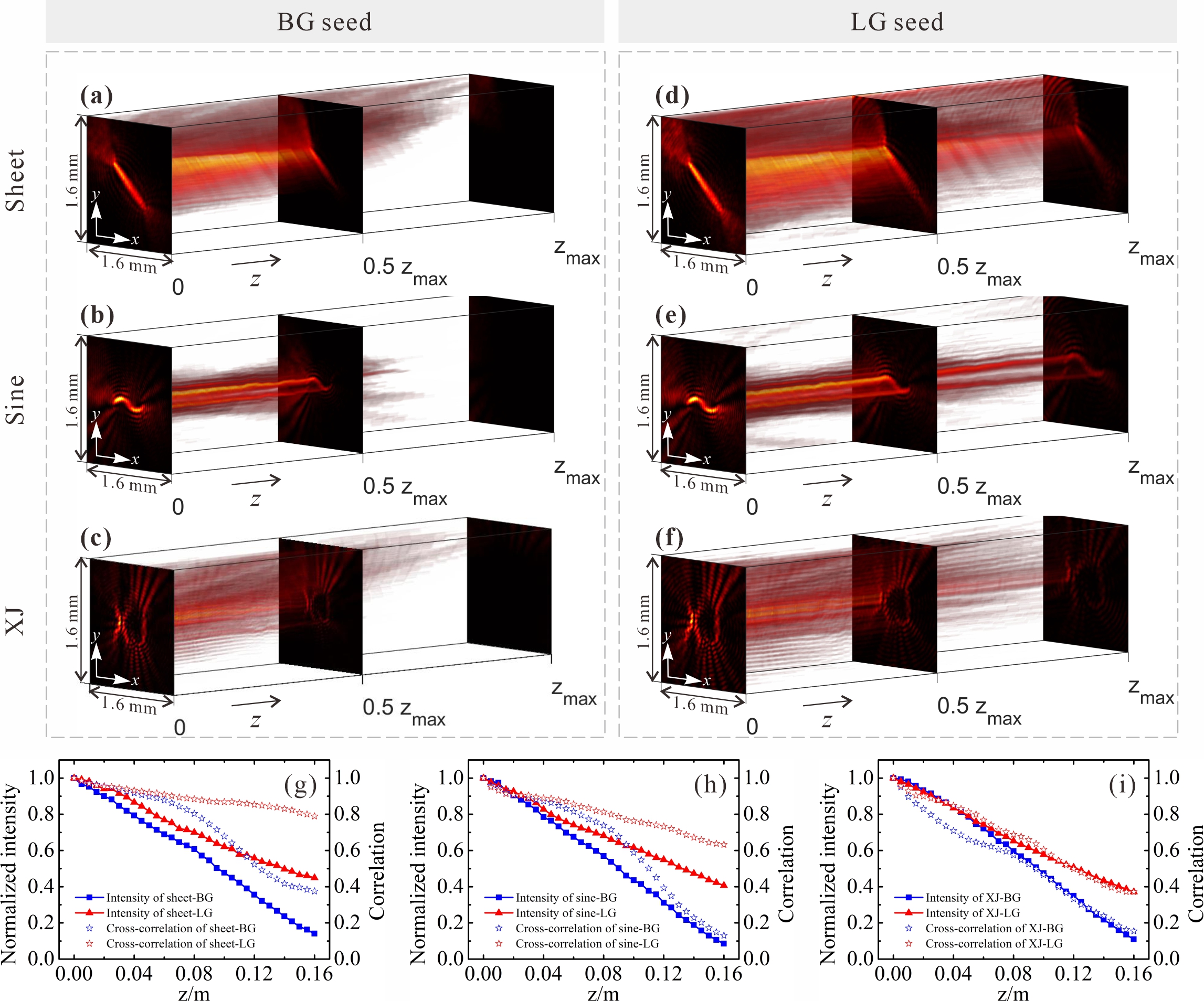}    %图片写法及名称格式
\\[1mm]    %图片与图说之间的距离，可修改，但应统一
\caption{
Experimental obtained normalized intensity volume for \textbf{(a)}-\textbf{(c)} customized-BG and \textbf{(d)}-\textbf{(f)} customized-LG beams. \textbf{(g)}-\textbf{(i)} Intensity and cross-correlation function variations against $z$ of the customized-BG and customized-LG beams.)
\hfill{}
}\label{fig:3}
%\vspace*{0.5mm}    %图片下空0.5mm
\end{figure*}

Next, the propagation of the three customized-LG/BG beams exhibited in Fig. \ref{fig:1} are studied experimentally in comparison. As shown by intensity volume in Fig. \ref{fig:3}, the transverse intensities of the customized-BG beams decrease dramatically during propagation and are almost invisible at $z=z_{max}$. In contrast, the intensities of the customized-LG beams attenuate much slower and are still recognizable at the farthest distance. This result can be easily inferred from the conclusion obtained in Fig. \ref{fig:2}, and is basically consistent with the theoretical simulation given in the supplementary. In addition, two noticeable phenomena are observed from Fig. \ref{fig:3}. The first is the non-uniform attenuation of the customized beams during their propagation, even for the symmetrically distributed sheet beams. This is mainly because the transverse intensity and phase structure of the customized beams, ``drawn'' by the two seed beams, are not exactly symmetric about the origin, and the superimposed seed beams will inevitably interfere with each other during propagation. The second is that the diffraction of LG leads to slight distortion of the customized-LG beam, especially for the beams with complex structure (see Fig. \ref{fig:3}(f)). Nevertheless, this shape change does not significantly alter its original profile, and more importantly its overall intensity is well maintained compared to the customized-BG beam.

For quantitative analysis, we extract the intensity summation of each pattern and plot the total intensity variation against $z$, as shown in Fig. \ref{fig:3}(g)-(i). One can find that in all cases given, the intensity of customized-LG is much higer than that of customized-BG beam after long-distance propagation. At $z=z_{max}$, the former still retains about 40\% of its original while the latter is reduced to about 10\%. Moreover, to assess the similarity between the transverse intensity distribution propagating to distance $z$ (labeled as $I_{z}$) and the original transverse intensity distribution $I_{0}$, we use the normalized two-dimensional cross-correlation function  which is simply defined as
\begin{equation}\label{Eq:7}
\gamma\left(z\right)=\frac{I_0 \star I(z)}{\sigma_{I 0} \sigma_{I (z)}},
\end{equation}
%\gamma\left(\mathbf{r}_{\perp}\right)=\frac{\left[I_0 \star I(z)\right]\left(\mathbf{r}_{\perp}\right)}{\sigma_{I 0} \sigma_{I (z)}}$\mathbf{r}_{\perp}$ refers to the transverse section,
where $I_0 \star I(z)$ denotes the cross-correlation function between $I_0$ and $I(z)$,  $\sigma_{I 0}$ and $\sigma_{I (z)}$ are standard deviations of the respective intensities. The maximum value of $\gamma(z)$ in each longitudinal position $z$ is shown with an asterisk in Fig. \ref{fig:3}(g)-(i) to represent the degree of correlation between $I_{z}$ and $I_{0}$. Evidently, the correlation has a maximum value of 1 at $z=0$ and decreases with the increase of $z$ due to non-uniform transverse intensity attenuation and subtle distortion of the light field structure. For the light with simpler designed shape, e.g., `sheet' and `sinusoid' shaped beams, this decrease occurs more slowly, as can be seen by comparing Fig. \ref{fig:3}(g) and (h), implying that they have a better performance in propagation invariability. On the other hand, for a specific target shape, the customized-LG beam has a higher correlation value than that of the customized-BG beam, especially at $z=z_{max}$, where the correlation difference between the two is up to about 50\%. Note that this difference is mainly due to the fact that the transverse intensity of the customized-LG decays more uniformly and slowly, but on the other side, the shape distortion of the complex structured beam (such as LG-XJ) caused by the diffraction of LG will reduce this correlation gap, as can be found in Fig. \ref{fig:3}(i). Even so, on the whole, the correlation value for customized-LG beam drops much slower than the customized-BG beam, especially over a distance of 0.5$z_{max}$. 

\begin{figure}[tb]    %此为跨栏图片写法，本杂志通栏宽度180mm,单栏宽度88mm，栏间距4mm
\vspace*{1mm}    %图片上空1mm
\centering    %图片居中
\includegraphics[width=8cm]{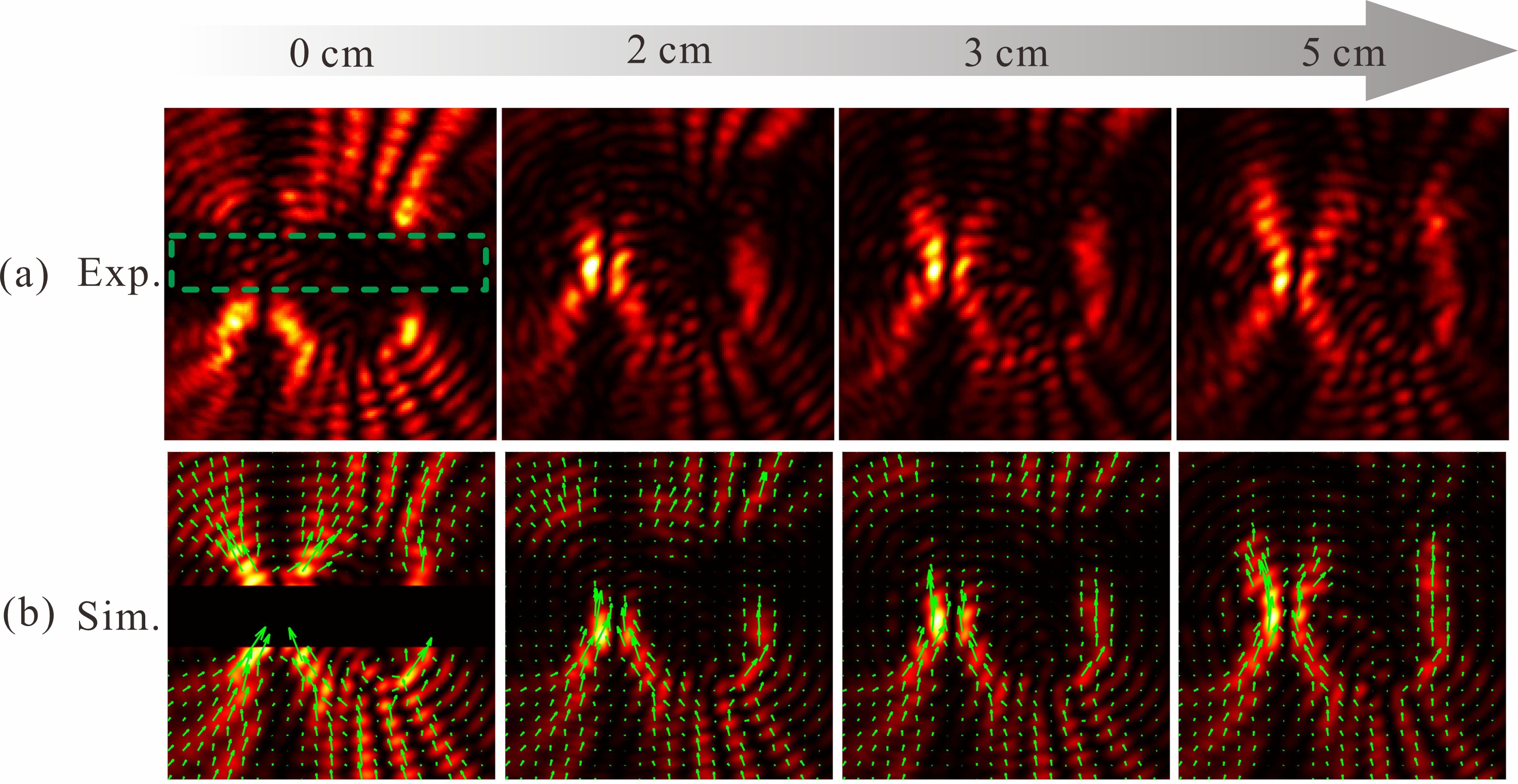}    %图片写法及名称格式
\\[1mm]    %图片与图说之间的距离，可修改，但应统一
\caption{
Self-healing process of the `XJ' shaped customized-LG beam. The first row is the experiment result and the second row is the corresponding theoretical simulation.)
\hfill{}
}\label{fig:4}
%\vspace*{0.5mm}    %图片下空0.5mm
\end{figure}

Finally, since it has been proved that the multi-ringed LG beam possesses the self-healing characteristic, here we further study whether the customized-LG beams heritage this characteristic. To this end, an obstruct is used to block parts of LG-XJ beam at $z=0$. The evolution of the blocked beam at different propagation distances is displayed in Fig. \ref{fig:4}(a). It is obvious that the occluded parts are almost restored at a propagation distance of 4 cm. The corresponding theoretical simulation, shown in Fig. \ref{fig:4}(b), agrees well with the experimental result. The transverse energy flow (indicated by white arrows) is also calculated by introducing the time-averaged Poynting vector under the paraxial approximation
\begin{equation}\label{Eq:8}
\langle\textbf{S}\rangle=\frac{i}{2}ck\epsilon_{0}(\psi^{*}\nabla\psi-\psi\nabla\psi^{*}),
\end{equation}
where $c$ is the speed of light in vacuum, $\epsilon_{0}$ is the vacuum permittivity, $\psi$ denotes the transverse light field and $*$ represents the complex conjugation item. The calculation manifests that the energy flux of the beam always flows along the customized curve. Consequently, the energy of the unoccluded region will continuously fill the occluded region and enable the beam to heal itself. Such an intriguing property may promote the application of customized-LG beams in a variety of complex environments.

\noindent
%%%%%%%%%%%%%%%%%%%%%%%%%%%%%%%%%%%%%%%%
%%%===============================================%%%

\section{Conclusion}
In summary, we experimentally generate a series of customized-LG beams that can present any desired high-intensity distributions. The key ingredient is based on tailoring the angular spectrum of the LG beam by an algebraic function, which is constructed using the Bessel-pencil method. We also generate some non-diffraction caustic beams that possess the same intensity profiles with the customized-LG beams for comparison. The results show that the customized-LG beams can maintain their profiles and suffer less energy loss than the non-diffraction caustic beams and hence can propagate a longer distance. Moreover, the self-healing ability of the customized-LG beam is further verified, which agrees well with the theory and validates the excellent performance of the customized-LG beam during propagation. This new phyletic structure beam could find broad applications in areas such as optical communication, spatial soliton control, optical tweezing and trapping, etc.

\section*{Funding}
This work is supported by National Natural Science Foundation of China (NSFC) (12104358,12104361, and 92050103) and Shaanxi Fundamental Science Research Project for Mathematics and Physics (22JSZ004). 

\bibliography{bib2}

%------------------------------------------------------------
\newpage 
\appendix

\onecolumngrid
\section{Experimental setup}

\begin{figure}[!h]
\centering\includegraphics[width=12cm]{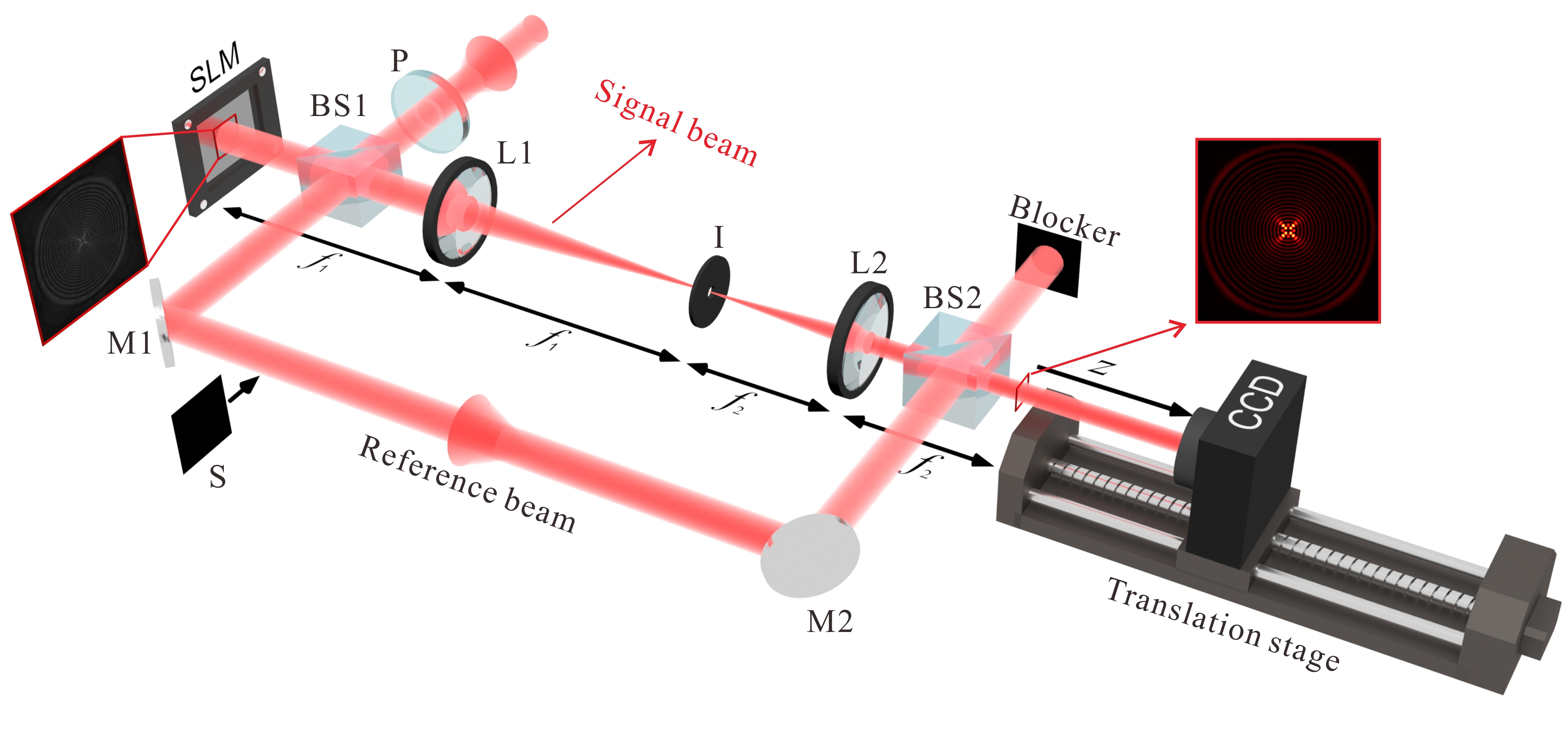}
\caption{Experimental setup. P: polarizer; L1-L2: lenses; M1-M2: mirrors; BS1-BS2: beam splitters; SLM: spatial light modulator; S: shutter; CCD: charge-coupled device.}\label{fig:s1}
\end{figure}

%In the following we experimentally generate several representative customized-PLG and customized-Bessel beams based on the methods above. 
The experimental setup is illustrated in Fig. \ref{fig:s1}. A horizontally polarized Gaussian beam (780 nm, Toptica DL) is perpendicularly incident on an SLM (HDSLM80R-PLUS) loaded with a pre-calculated hologram carrying the complex amplitude information of the target light field $\mathcal{A}\tilde{U}_{0}$. The beam diffracted from SLM is then spatially filtered by a filtering system composed of two lenses (with focal lengths of f1 = 250 mm and f2 = 150 mm respectively) and an iris, and its transverse intensity is recorded by a CCD (Lumenera INFINITY 3). The position of the CCD can also be shifted through a translation stage to record the intensity patterns at different propagation distances. The intensity patterns can be synthesized into a three-dimensional intensity volume, with an additional dimension representing the propagation distance. Each intensity volume is normalized by its maximum value, as shown in Fig. 3(a)-(f) in the manuscript. During the intensity recording process, it is always ensured that the CCD is not overexposed and the exposure time and gain are maintained, and each intensity pattern is averaged 5 times to minimize random errors. Besides, an additional reference beam can be switched on by a shutter (S) to interfere with the customized beams for phase measurements. 

%%%===============================================%%%
%%%===============================================%%%

\section{Intensity and phase of experimentally customized light fields}

%%%===============================================%%%

\begin{figure}[!h]
\centering\includegraphics[width=11cm]{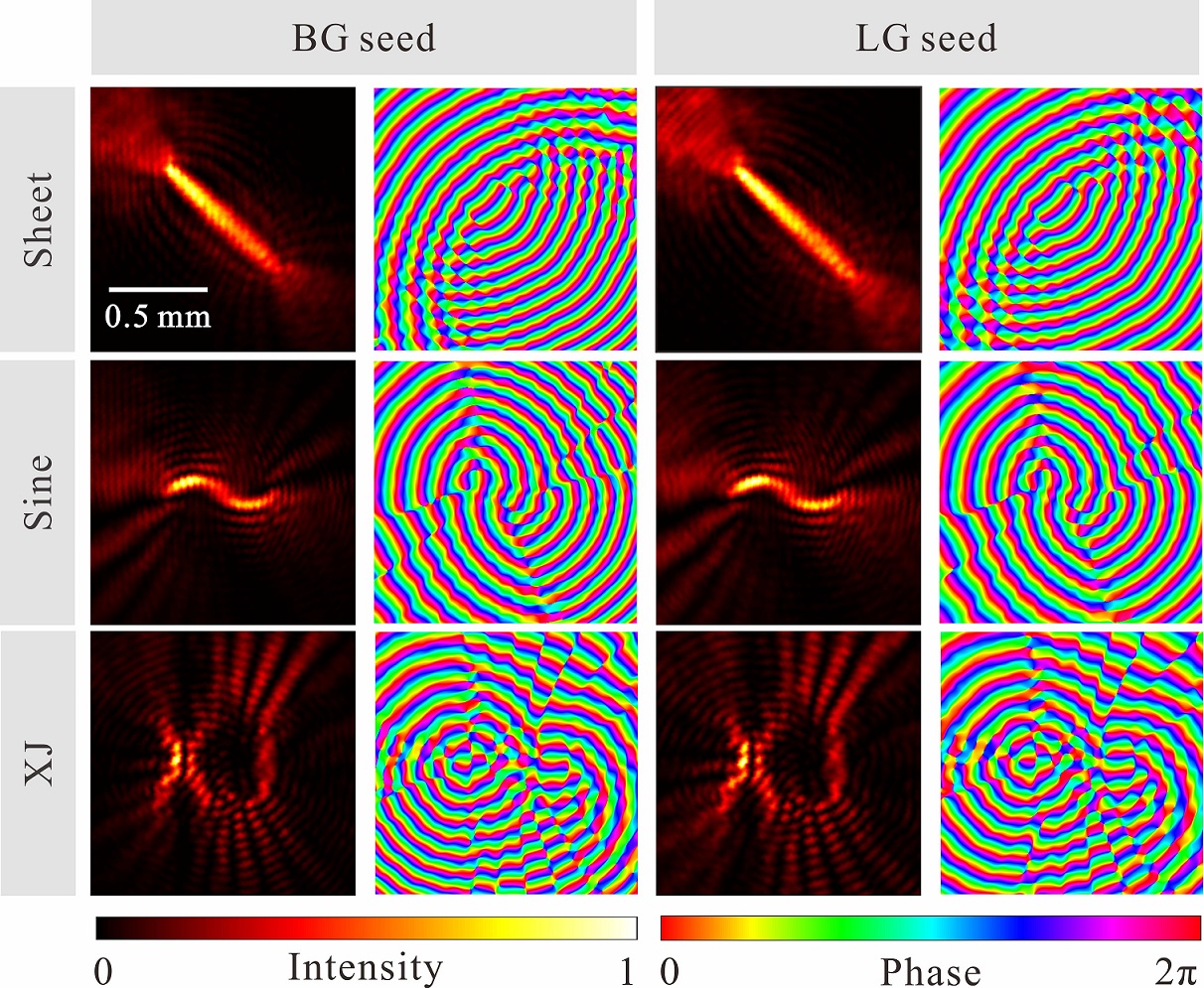}
\caption{Experimentally measured transverse intensities and phases of the customized beams.}\label{fig:s2}
\end{figure}

We use a reference beam to interfere with the customized beams and record the interference patterns, then apply the holographic interferometry method \cite{2015Analog} to determine the phase distributions of the customized beams, as shown in the second and fourth columns of Fig. \ref{fig:s2}. The intensity profiles of the object beams are also reconstructed for comparison, as shown in the first and third columns of Fig. \ref{fig:s2}. All the experimental and theoretical (see Fig. 1(b)) results are in good agreement, indicating that the qualities of the customized beams are very well.

\begin{figure}[htp]
\centering\includegraphics[width=13cm]{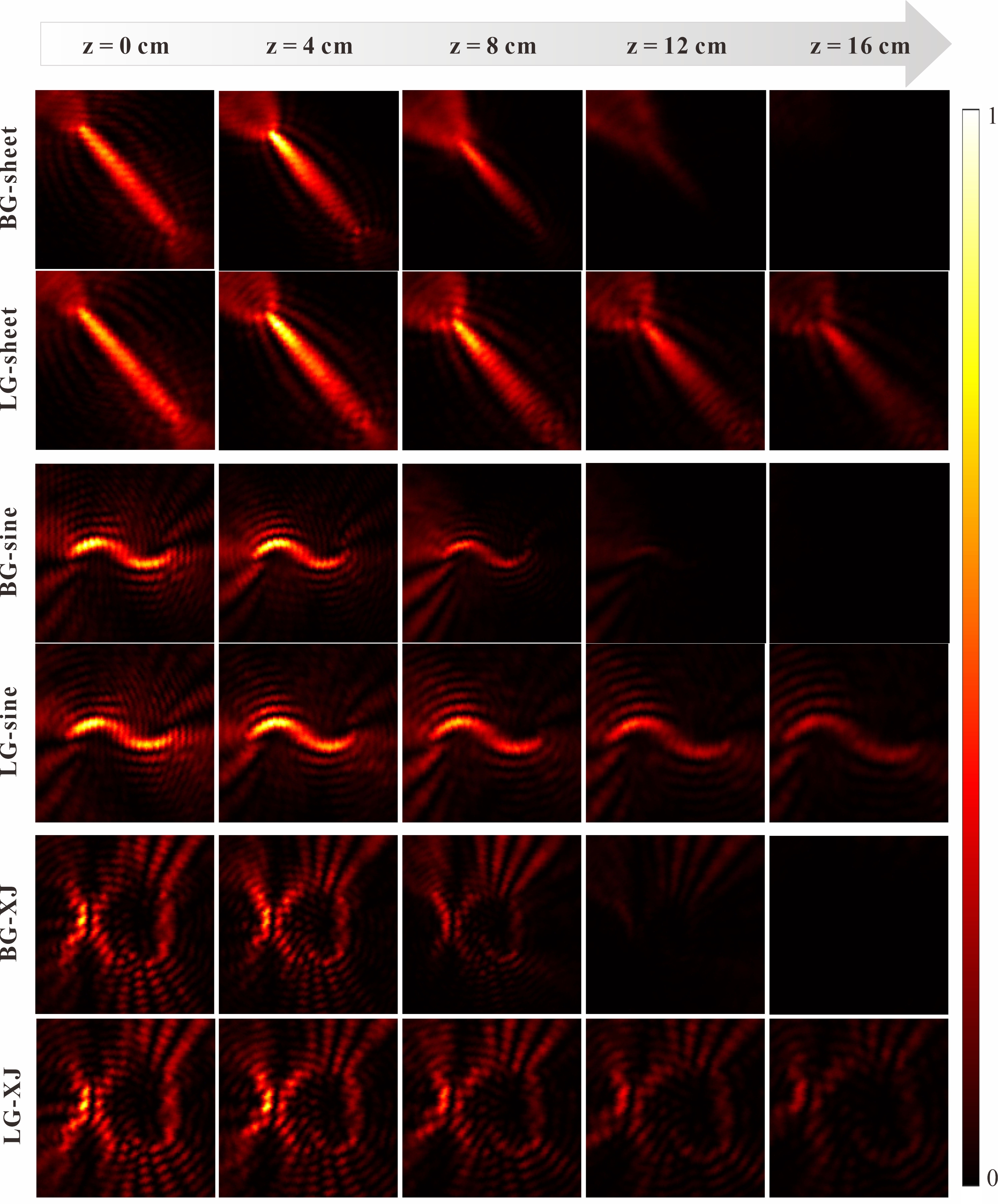}
\caption{Normalized intensity distribution in the $x-y$ plane of customized light fields at various propagation distance $z$.}\label{fig:s5}
\end{figure}

Fig. 3(a)-(f) illustrates the 3-dimensional normalized intensity volume of the customized beams. Here we further give the transverse intensities of the corresponding beams at 5 different propagation distances, as shown in Fig. \ref{fig:s5}.  

%%%===============================================%

\noindent

\section{Propagation of astroid-BG/LG and deltoid-BG/LG beams}
\begin{figure}[!h]
\centering\includegraphics[width=12cm]{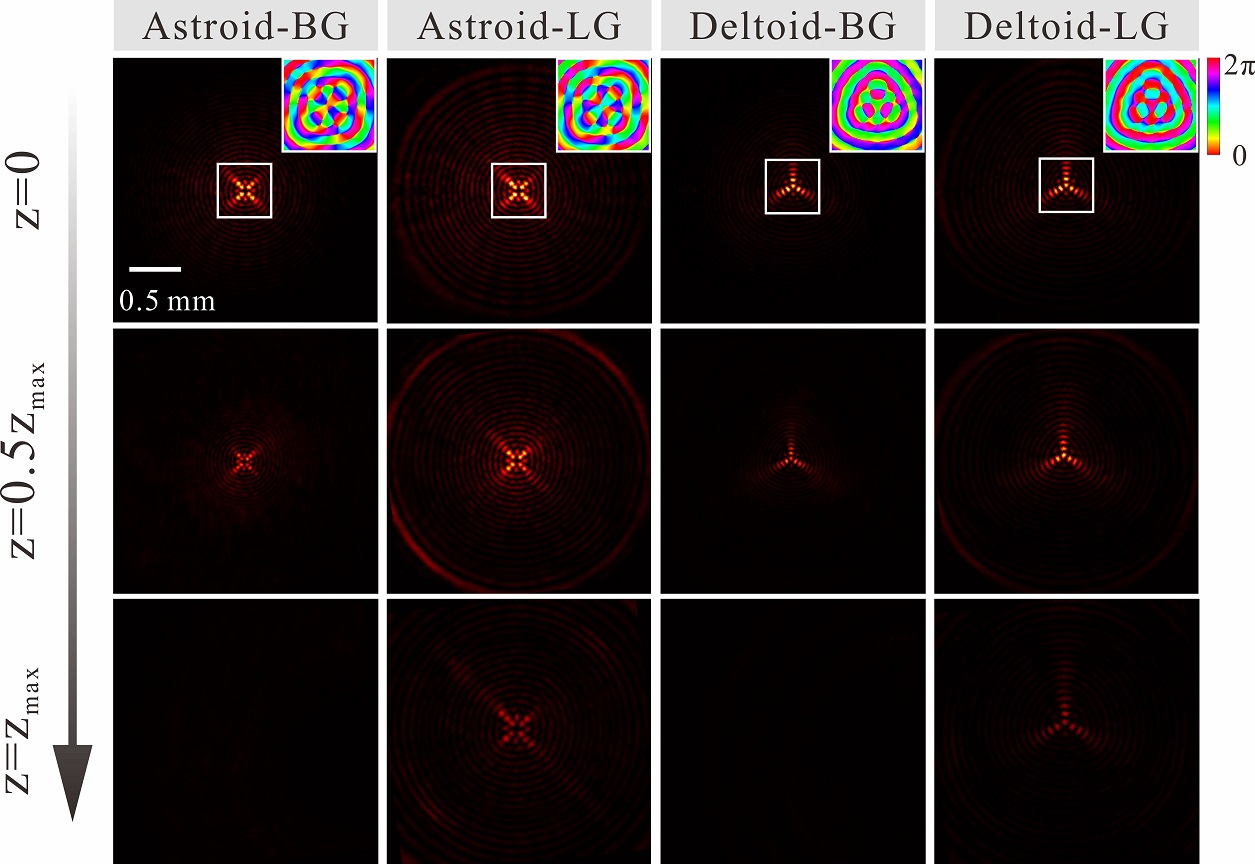}
\caption{Transverse intensity of the customized beams at different propagation distances. The inset reveals the phase corresponding to the region indicated by the white box.}\label{fig:s3}
\end{figure}

Figure \ref{fig:s3} shows the transverse intensity of astroid-BG/LG and deltoid-BG/LG beams propagating to 0, 0.5 $z_{max}$, $z_{max}$. Here, the spectrum modulation is set to $\mathcal{A}=\exp[i(5\sin(2\varphi)/2)]$ for the astroid-BG/LG beams and $\mathcal{A}=\exp[i5\sin(\varphi)^3]$ for the deltoid-BG/LG beams. It can be seen that similar to the results revealed in Fig. 3, the astroid-BG (or deltoid-BG) and astroid-LG (or deltoid-LG) initially have very similar intensity distributions due to undergoing the same spectrum modulation. But after transmission to the farthest distance, the latter still maintains clear shapes, although its outline is slightly expanded, while the former almost disappears in the field of view.

%%%===============================================%%%
\section{Theoretical simulations of the customized-BG/LG beams propagation in free space}

%To make the experimental results more convincing,
The customized beam is obtained by acting a linear differential operator $\hat{\mathrm{A}}\left(D_x, D_y\right)$ on a seed beam $U_{0}(r,z)$. $\hat{\mathrm{A}}\left(D_x, D_y\right)$ can be expressed as a sum of products of derivatives 
\begin{equation}\label{Eq:1}
\hat{\mathrm{A}}\left(D_x, D_y\right) \equiv C_{a, b} D_x^a D_y^b,
\end{equation}
where $D_x^a$ and $D_y^b$ are the $a$th-order and the $b$th-order derivatives with respect to $x$ and $y$, $C_{a, b}$ is amplitude coefficient. $U_{0}(r,z)$ is a solution of the paraxial wave equation (PWE) and $\hat{\mathrm{A}}\left(D_x, D_y\right)$ commutates with the operator of the PWE. So $U(\mathbf{r},z)=\hat{\mathcal{A}}U_{0}(r,z)$ is a new solution of the PWE. Let $\mathcal{F}$ and $\mathcal{F}^{-1}$ represent the two-dimensional Fourier transform and inverse Fourier transform, and $U(\mathbf{r},z)=\hat{\mathcal{A}}U_{0}(r,z)$ can be rewritten in Fourier space as 
\begin{equation}\label{Eq:2}
U=\mathcal{F}^{-1}\left\{\mathcal{F}\left\{\hat{\mathcal{A}} U_{0}\right\}\right\}=\mathcal{F}^{-1}\left\{\mathcal{A} \mathcal{F}\left\{U_{0}\right\}\right\}=\mathcal{F}^{-1}\left\{\mathcal{A} \tilde{U}_{0}\right\},
\end{equation}
where $\tilde{U}_{0}$ is the Fourier transforms $U_{0}$, and $\mathcal{A}$ is an algebraic function that represents the operator $\hat{\mathcal{A}}$ on the Fourier plane. A typical example related to $\hat{\mathcal{A}}$ is the ladder operator $\hat{\mathcal{L}}^{\pm}=\partial_{x}\pm i\partial_{y}$, with its algebraic function being $\mathcal{L}^{\pm}=\mathcal{F}[\hat{\mathcal{L}}^{\pm}]=-ik_{x}\pm k_{y}\propto k_{r} \exp(\pm i\phi)$, where ($k_{x}, k_{y}$) = ($k_{r} \cos \phi, k_{r} \sin \phi$) represents Fourier coordinates. It can be verified that the operator $\hat{\mathcal{L}}^{\pm}$, when applied to the 0th-order Bessel beam with wavenumber $k_{t}$, yields the 1st-order Bessel beam, i.e., $\hat{\mathcal{L}}^{\pm}J_{0}(k_{t}r)=\mathcal{F}^{-1}[\mathcal{L}^{\pm}\mathcal{F}[J_{0}(k_{t} r)]]\propto\mathcal{F}^{-1}[\delta (k_{r}-k_{t}) \exp(\pm i\phi)]\sim J_{1}(k_{t} r)\exp(\pm i\phi)$.

An arbitrary field $U(\textbf{r})$ with diffraction-free properties can be expressed using Whittaker's integration
\begin{equation}
\label{eq.S6}
U(\textbf{r})=\oint\mathcal{A}(\varphi)\exp [ik_{t}\textbf{r}\cdot \textbf{u}(\varphi)]d\varphi,
\end{equation}
where $\mathcal{A}(\varphi)$ is the algebraic function employed to modulate the shape of $U(\textbf{r})$, $k_{t}$ represents the transverse wavenumber, $\textbf{u}(\varphi)=(\cos \varphi,\sin \varphi)$ is a unit vector related to the azimuthal angle. When $\mathcal{A}=1$, $U(\textbf{r})$ corresponds to the 0th-order Bessel beam (denoted as $U_{B0}$) and locates at the origin of coordinates. If we aim to shift the center of $U_{B0}$ to the position $\textbf{r}=\textbf{r}_{c}$, $\textbf{r}$ in Eq. (6) should be replaced by $(\textbf{r}-\textbf{r}_{c})$. The optical field after this movement is represented by
\begin{equation}
\label{eq.S7}
U_{B0}(\textbf{r}-\textbf{r}_{c})=\oint\exp [ik_{t}(\textbf{r}-\textbf{r}_{c})\cdot \textbf{u}(\varphi)]d\varphi.
\end{equation}

Comparing Eq. (6) and Eq. (7), we can observe that the algebraic function $\mathcal{A}$, corresponding to the 0th-order Bessel beam with the field center located at $\textbf{r}=\textbf{r}_{c}$, is $\mathcal{A}=\exp[-ik_{t}\textbf{r}_{c}\cdot\textbf{u}(\varphi)]$. When we use $U_{B0}$ as a `pencil' and continuously shift it along a preset curve $\textbf{r}_{c}(\tau)$, then the corresponding $\mathcal{A}$ should be integrated over the curve, i.e., $\int_{\textbf{r}_{c}(\tau)}\exp[-ik_{t}\textbf{r}_{c}\cdot\textbf{u}(\varphi)]d\tau$. However, it should be mentioned that the superposition of the 0th-order Bessel beam along the aforementioned curve may impact the shape of the customized beam, which is undesirable. Hence, it is necessary to introduce an additional phase term $\gamma_{B}$ into the integral. By utilizing the generalized law of refraction \cite{doi:10.1126/science.1210713}, it can be established that for a customized beam using a pre-defined curve $\textbf{r}_{c}(\tau)$, its intensity profile $r_{c}$ satisfies $r _{c}= \frac{1}{k_{t}}\frac{d \arg[\mathcal{A}(\varphi)]}{d\varphi}$, where $\arg[x]$ is the angle phase of $x$. Consequently, it can be inferred from a simple example, such as when $\textbf{r}_{c}$ represents a circular ring, resulting in $\arg [A]=2\pi r_{c} k_{t}=k_{t}\oint|\textbf{r}_{c}(s)|ds$, that $\gamma_B$ should be proportional to the arc length of the curve $\textbf{r}_{c}$: $\gamma_{B}(\tau)=k_{t}\int_{0}^{\tau}|\textbf{r}_{c}(s)|ds $. Incorporating all the above analyses, we can obtain the mathematical description of $\mathcal{A}$ corresponding to the target curve $\textbf{r}_{c}(\tau)$
\begin{equation}
\label{eq.S8}
\mathcal{A}(\varphi)=\int_{\textbf{r}_{c}}\exp[i\gamma_{B}(\tau)-ik_{t}\textbf{r}_{c}\cdot\textbf{u}(\varphi)]d\tau.
\end{equation}

\begin{figure}[!h]
\centering\includegraphics[width=\linewidth]{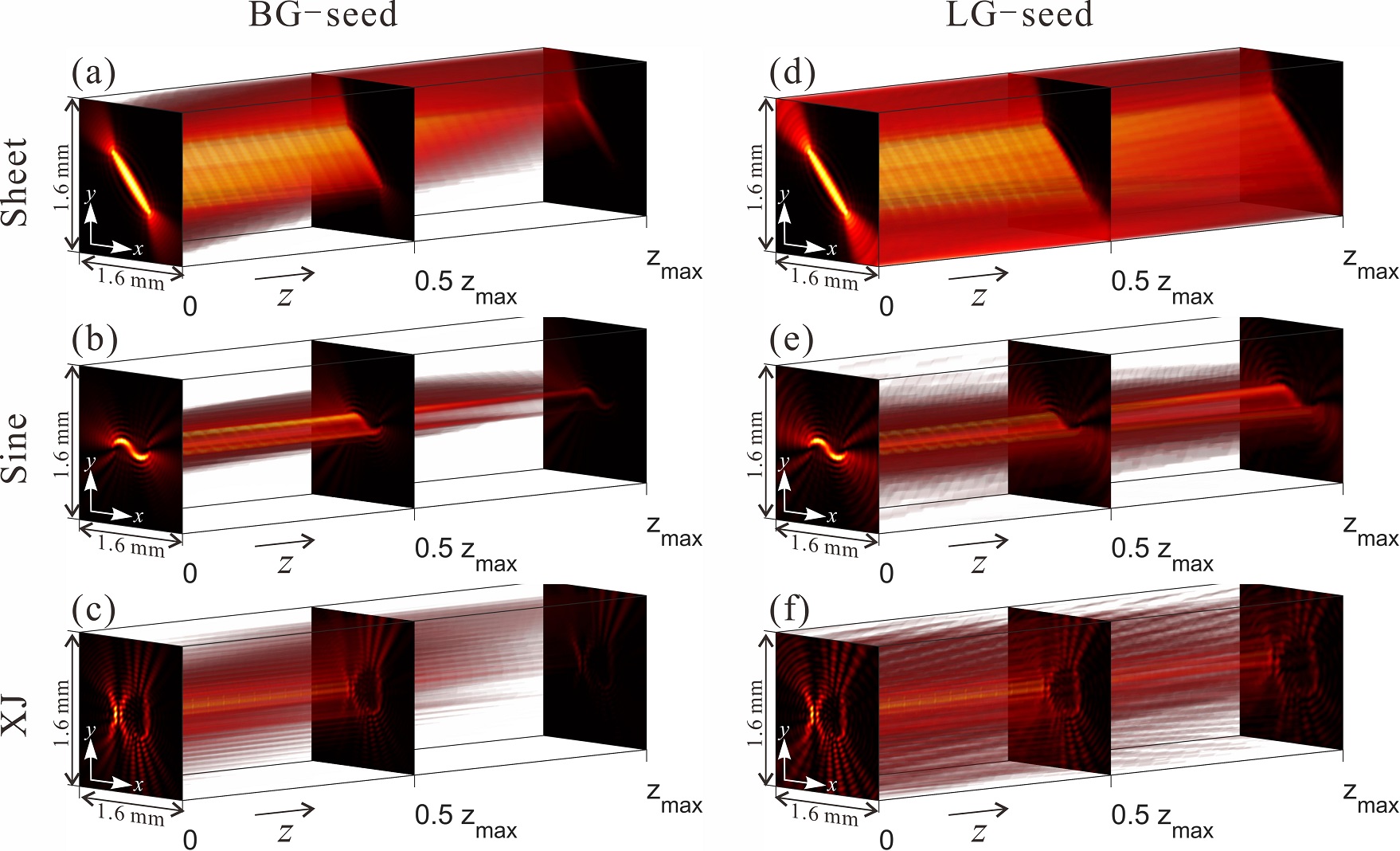}
\caption{Theoretical normalized intensity volume for (a)-(c) customized-BG and (d)-(f) customized-LG beams}\label{fig:s4}
\end{figure}

We use the theory above, as well as the scalar diffraction theory and fast Fourier transform algorithm \cite{voelz2011computational} to simulate the propagation of the customized beams shown in Fig. 3. The calculated normalized intensity volume is displayed in Fig. \ref{fig:s4}, from which we can see that compared with the customized-LG beams, the customized-BG with similar spatial structure experienced more serious intensity attenuation during propagation, which is consistent with the experimental results given in Fig. 3.

We also give a qualitative analysis of why the intensity of the customized-LG beams decays slower than that of the customized-BG beams, as detailed below. The customized optical field can be regarded as an ordered superposition of the seed beams. Therefore, the intensity attenuation rate of the customized beam is similar to that of the seed beams. Given that the intensity of the seed beam is highly localized to the center point, the intensity decay phenomenon can be comprehended by analyzing the on-axis intensity variation with the propagation distance of the seed beam. 
From the general expression of LG beams, the on-axis field of an LG seed (when $l=0$) satisfies
\begin{equation}
\label{eq.S9}
LGU_{0}(r=0,z)\propto\frac{1}{\omega_{0} \sqrt{1+(z/z_{r})^{2}}}L_{p}^{0}(0),
\end{equation}
\noindent where $z_{r} = \pi \lambda \omega_{0}^{2}$ is the Rayleigh length, equivalent to $z_{max}$ as mentioned in our main text. From Eq. (\ref{eq.S9}), we can further derive the on-axis intensity of the LG seed as
\begin{equation}
\label{eq.S10}
I_{LG}(r=0,z)\propto\frac{1}{ 1+(z/z_{r})^{2}}.
\end{equation}

Regarding the BG seed beams described by Eq. (4) in the manuscript, its amplitude of the electric field evolving with propagation distance $z$ is typically given by \cite{GORI1987491}
\begin{equation}
\label{eq.S11}
|BU_{0}(r,z)|= A\frac{\omega_{B}}{\omega(z)}J_{0}\left[k_{t}r/(1+iz/z_{Br})\right]\exp\left[-r^{2}/\omega^{2}(z)+k_{t}^{2}z^{2}/k^{2}\right],
\end{equation}
\noindent where $z_{Br} = \pi \lambda \omega_{B}^{2}$ and $\omega(z)= \omega_{B}\sqrt{1+(z/z_{Br})^2}$. Hence, we obtain the on-axis intensity of the BG seed as
\begin{equation}
\label{eq.S12}
I_{BG}(r=0,z)=|BU_{0}(r=0,z)|^{2}\propto \frac{1}{1+(z/z_{Br})^2}\left[\exp\left(-\frac{k_{t}^{2}z^{2}}{k^{2}\omega^{2}(z)}\right)\right]^{2}.
\end{equation}

Considering that $z$ is within the range of $0$ to $z_{r}$ (or $z_{max}$) and $\omega_{B}=\sqrt{2N}\omega_{0}\gg \omega_{0}$ (since $N=p+1/2\gg 1$), it follows that $z/z_{Br}\leq z_{r}/z_{Br}=(\omega_{0}/\omega_{B})^{2}\ll 1$, thereby resulting in $1+(z/z_{Br})^{2}\approx 1$ and $\omega(z) \approx \omega_{B}$. Incorporating these approximations with the relationship of $k_{t}=2\sqrt{2N}/\omega_{0}$ and $\omega_{B}=\sqrt{2N}\omega_{0}$, Eq. (\ref{eq.S12}) can be simplified to the form:
\begin{equation}
\label{eq.S13}
I_{BG}(r=0,z)\propto \exp(-z^{2}/z_{r}^{2}).
\end{equation}
\noindent

\begin{figure}[tp]
\centering\includegraphics[width=\linewidth]{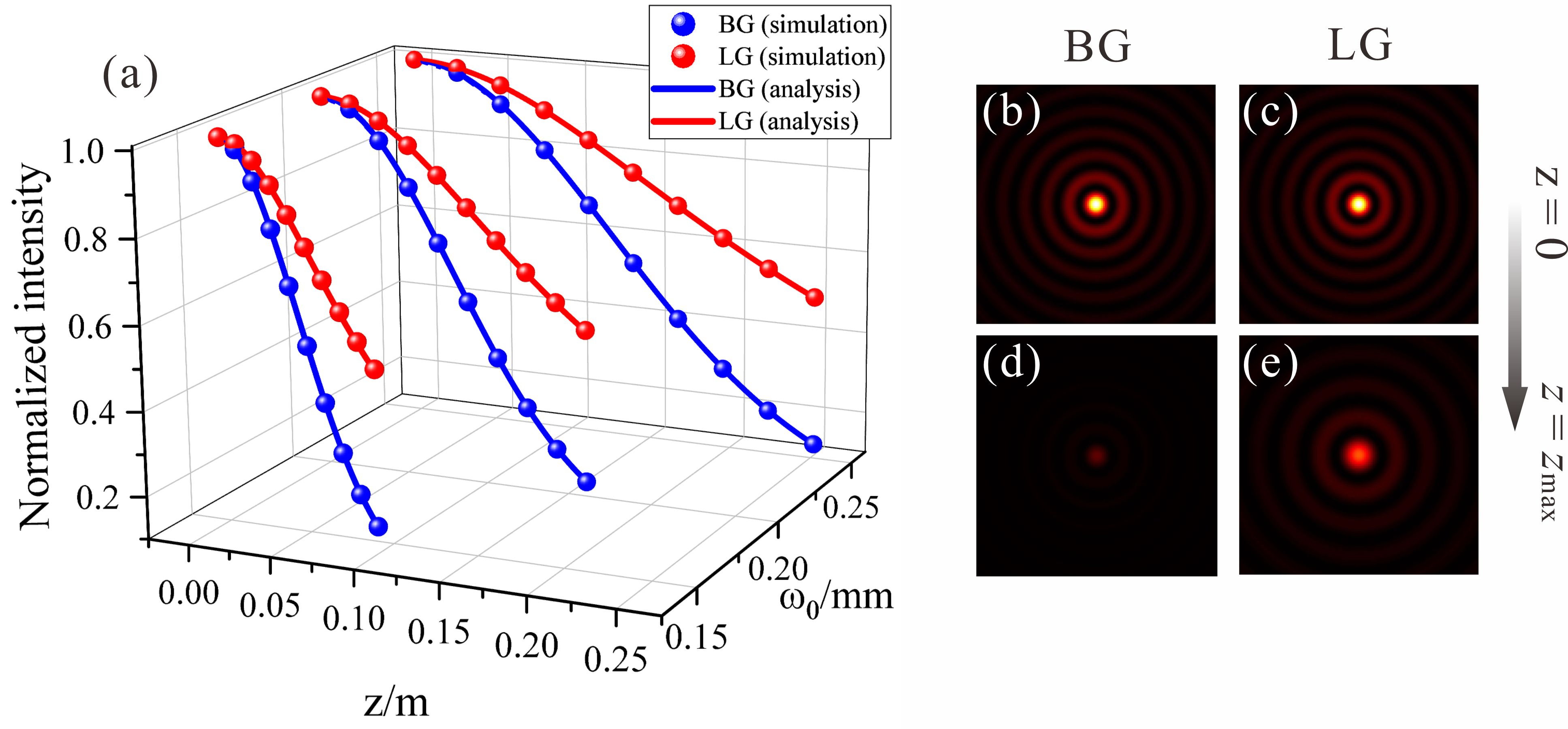}
\caption{(a) The on-axis intensity variation with propagation distance $z$ for two seed beams under different waist sizes $\omega_{0}$. (b)-(e) Normalized intensity of BG and LG beams with $\omega_{0}=0.25$ mm at $z=0$ [(b) and (c)] and $z=z_{max}$ [(d) and (e)]. The simulations are performed with the parameter set to $p=20$ and $\lambda=780$ nm.}\label{fig:s7}
\end{figure}

Comparing Eq. (\ref{eq.S10}) and Eq. (\ref{eq.S13}), we can see that $I_{LG}(r=0,z)\geq I_{BG}(r=0,z)$ consistently holds for $0\leq z\leq z_{r}$. This means that the intensity attenuation rate of the LG seed is slower than that of the BG seed, regardless of the values of $\omega_{0}$ and $p$. Using the analytical formula (i.e., Eq. (\ref{eq.S9}) and (\ref{eq.S12})) and the angular spectrum propagation method, we obtain the variations of $I_{BG}(r=0,z)$ and $I_{LG}(r=0,z)$ with the propagation distance under different $\omega_{0}$, as shown in Fig. (\ref{fig:s7}). This trend is consistent with that of the customized beam in the manuscript.

\noindent

%\bibliography{sample}

\end{document}